\documentclass[aps,preprint,nofootinbib,preprintnumbers,eqsecnum,superscriptaddress]{revtex4}
\pdfoutput=1

\usepackage{wrapfig}

\usepackage{subfigure}

\usepackage{color}

\usepackage[
      colorlinks=true,
      linkcolor=blue,
      urlcolor=blue,
      filecolor=black,
      citecolor=red,
      pdfstartview=FitV,
      pdftitle={},
        pdfauthor={},
        pdfsubject={},
        pdfkeywords={},
        pdfpagemode=None,
        bookmarksopen=true
      ]{hyperref}

\usepackage[font=footnotesize,labelfont=bf,justification=centerlast,width=.94\textwidth]{caption}

\usepackage[normalem]{ulem}
\usepackage{amsmath}
\usepackage{enumerate}
\usepackage{amsfonts}
\usepackage{yfonts}

\usepackage{subfigure}
\usepackage{psfrag}

\usepackage{epsfig}
\usepackage[latin1]{inputenc}
\usepackage{float}
\usepackage{graphicx}
\usepackage{cancel}
\usepackage{mathrsfs}
\usepackage{amssymb}
\usepackage{amsfonts}
\usepackage{amsmath}
\usepackage{slashed}

\usepackage{graphicx}
\usepackage{bm}

\newcommand{\be}{\begin{equation}}
\newcommand{\ee}{\end{equation}}
\newcommand{\bea}{\begin{eqnarray}}
\newcommand{\eea}{\end{eqnarray}}
\begin{document}

\title {A little walk from physical to biological complexity: \\ protein folding and stability}

\author{Fabrizio Pucci, Marianne Rooman}
\affiliation{BioModeling, BioInformatics \& BioProcesses, Universit\'e Libre de Bruxelles,
CP 165/61, 50 Roosevelt Ave, 1050 Brussels, Belgium}

\bigskip 
\begin{abstract}
\vskip 0.3in
As an example of topic where biology and physics meet, we present the issue of protein folding and stability, and  the development of thermodynamics-based bioinformatics tools that predict the stability and thermal resistance of proteins and the change of these quantities upon amino acid substitutions. These methods are based on knowledge-driven statistical potentials, derived from experimental protein structures using the inverse Boltzmann law.  We also describe an application of these predictors, which contributed to the understanding of the  mechanisms of aggregation of a particular protein known to cause  a neuronal disease.

\begin{center}
{\it To Philippe Spindel for his 65th birthday}
\end{center}
\end{abstract}

\begin{figure}[b]
\begin{center}
\includegraphics[scale=0.5]{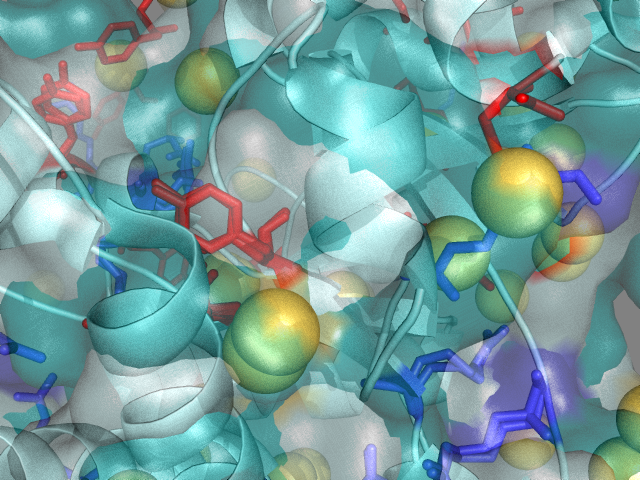}
\end{center}
\end{figure}

\maketitle

\vfill


\tableofcontents

\section{Introduction}

Molecular and cellular biology were traditionally purely experimental sciences, and it is only recently that theoretical biology -- including bioinformatics, biomodeling and biomathematics -- has started to develop.  The need for informatics approaches  has become essential with  the advent of high-throughput  technologies, which generate a huge amount of data in a single experiment, such as  full genome sequences or the ensemble of transcribed RNA molecules in a cell  population. These technologies are at the basis  of omics science, defined as the collective study and characterization of  biomolecules: proteomics is the large-scale study of proteins, genomics of genomes, transcriptomics of  transcribed RNA, mutatomics of all mutations in proteins or DNA, and so on.  

In parallel to omics science, (computational) systems biology has started to attract a lot of attention. It has become clear that the functioning of cells cannot be understood only through the study of individual macromolecules, such as proteins, RNA, and DNA. Indeed, the interactions between all biomolecules present inside and outside the cells are of prime importance, and they are  indispensable to understand and model cellular life. In other words, the division of the cell  into independent subsystems is a much too strong approximation. Obviously, this has the consequence of dramatically increasing  the complexity of the systems to be studied. 

An interesting feature of bioinformatics and biomodeling is that application is never far. Computational approaches are increasingly exploited to rationalize existing data and to develop knowledge-based predictors. These are  unable to give a unique optimal solution -- as the systems are far too complex --, but propose valuable candidate hypotheses, which need to be further analyzed and validated by experimental means. These  targeted conjectures save a lot of time as they allow focusing on a few relevant possibilities. In the medical sector,  computational approches are more and more used in rational drug \cite{drugdesign}, antibody \cite{antibody} and vaccine \cite{vaccine} design, or for identifying genome variants that cause human diseases \cite{SNP1,SNP2}. In the very near future, therapies will be personalized for each patient on the basis of the variants that appear in his genome. On the other hand, enzymes are used for their unique catalytic properties in many biotechnology sectors, and their optimization is an important challenge for setting up efficient and environment-friendly bioprocesses \cite{BiotechI,BiotechII,BiotechIII}. Finally, synthetic biology  will certainly gain importance in the  future, and modified or synthetic cells can be expected to be used  as little factories that produce specific compounds to be used for example as energy source ({\it e.g.} bioethanol, biohydrogen) \cite{synthetic1, synthetic2}.  

\begin{figure}[h]
\begin{center}
\subfigure[]
{\includegraphics[scale=0.4]{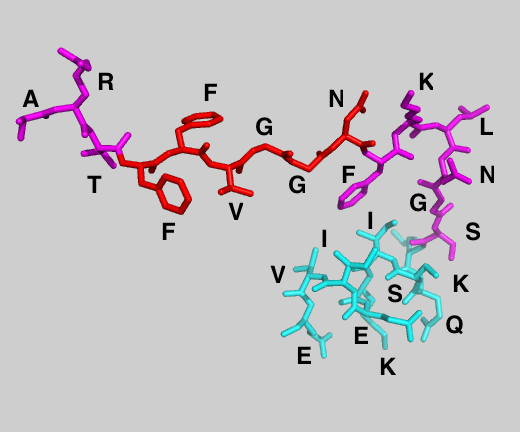}}
\subfigure[]
{\includegraphics[scale=0.4]{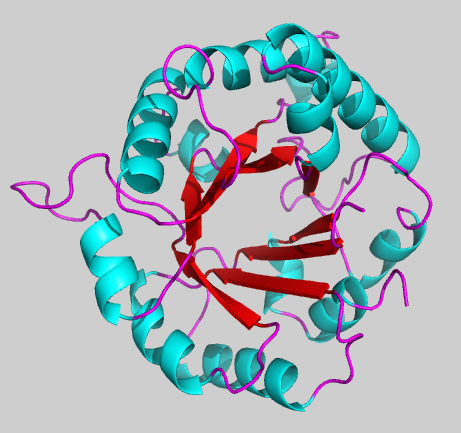}}
\caption{(a) Part of a protein chain. Proteins are sequences of about 50 to more than 10,000 covalently bound amino acids. These are of  20 types: the 20 amino acids that are encoded in the genome. They share a common part (called main chain) and differ by their  side chains which carry molecular groups with different physicochemical properties (charged, aromatic, aliphatic, etc). They are represented by uppercase letters. (b) Native (folded) structure of triose phosphate isomerase (Protein DataBank (PDB) \cite{PDBank} code: 7tim). Helices are represented in blue, $\beta$-strands in red and loops in magenta.} 
\label{prot}
\end{center}
\end{figure}

We focus in this paper on a specific subdomain of structural bioinformatics, {\it i.e.} protein folding and stability. Proteins are chains of amino acids and a large class of them fold  into a well-defined 3-dimensional (3D) structure in physiological conditions (Fig. \ref{prot}). Generally, when correctly folded, they fulfill their biological role (catalysis, molecular transport, signal transduction, genome regulation, etc). Protein folding is (usually) a reversible process basically ruled by classical physics, in which the 3D structure is stabilized by interactions between the different types of amino acids. The problem is thus well defined and can be tackled with physical methods, specifically through statistical mechanics. However, it is extremely difficult to actually predict the 3D structure from the amino acid sequence, for two reasons: the conformational space of a protein is enormous and  the energy functions that describe inter-residue interactions are not perfectly accurate. 

We present some bioinformatics tools that were developed in our group in view of estimating  the thermodynamic and thermal stability of a protein, and  the changes that occur upon substitution of one amino acid into another \cite{scoop1,scoop2,PoP1, PoP2,HoT}. Such  tools are very helpful in protein design, when aiming at rationally modifying or optimizing certain protein characteristics, whether in biopharmaceutical, biomedical or biotechnological applications, or simply when attempting to elucidate the complex relationships between protein sequence, structure, stability, interactions and function.

\section{Protein folding and stability}\label{sec:pfs}

In general, proteins occur in two distinct structural states in their usual environment, that is the intra- or extracellular medium, or the cellular membrane. These are the folded and unfolded state, which correspond to two free energy minima. The folded, biologically active, state has a low entropy but favorable enthalpy, while the unfolded state has a large entropy but unfavorable enthalpy. The (un)folding transition is an equilibrium process, where the folded structure is preferred for a range of conditions including the physiological conditions ({\it e.g.} the living temperature and pH of the host organism), and the unfolded structure is preferred for the other conditions. The  transition is usually reversible, at least if aggregation or post-translational modifications are overlooked and if large multidomain proteins are not considered. Note that this is true for the class of proteins called structured proteins.  Another class consists of proteins that are intrinsically unstructured or  only fold in the presence of some biomolecular partner; we will not consider such proteins here.

Proteins play crucial roles in and outside the cells -- they do most of the actual work required for cellular life. Mutations in proteins define polymorphisms -- which are at the basis of the phenotypic variability between individuals. However, though the large majority of  mutations have a neutral effect, some significantly affect the protein structure, stability, interactions or function, and cause  diseases.

\begin{figure}[h]
\begin{center}
\subfigure[]
{\includegraphics[scale=0.185]{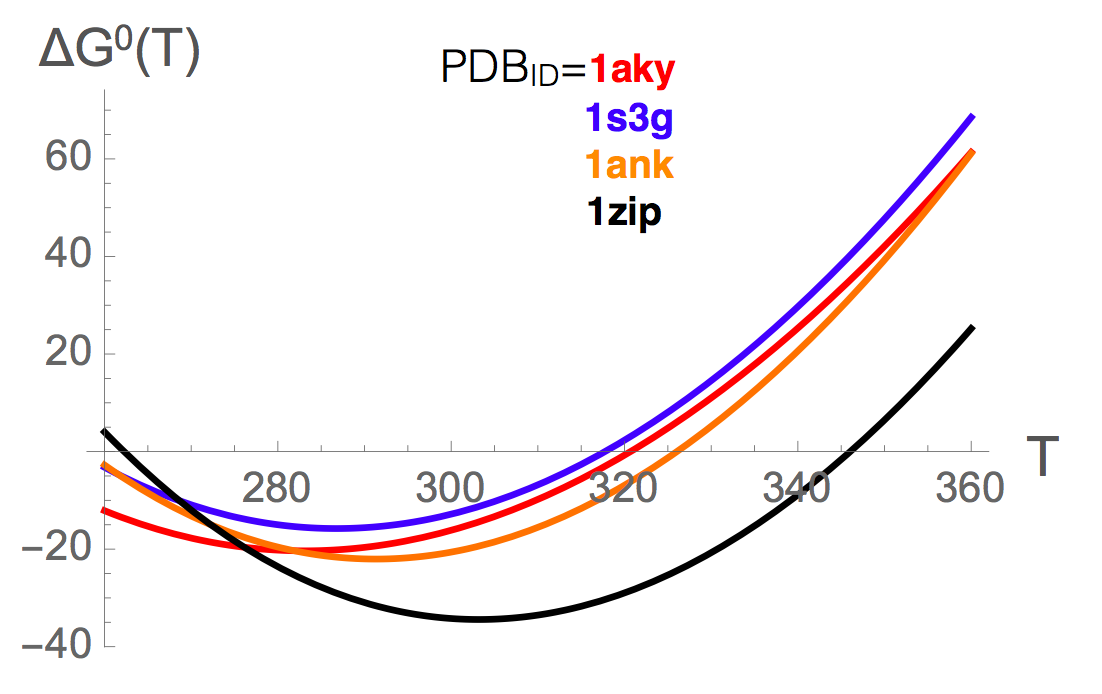}}
\subfigure[]
{\includegraphics[scale=0.185]{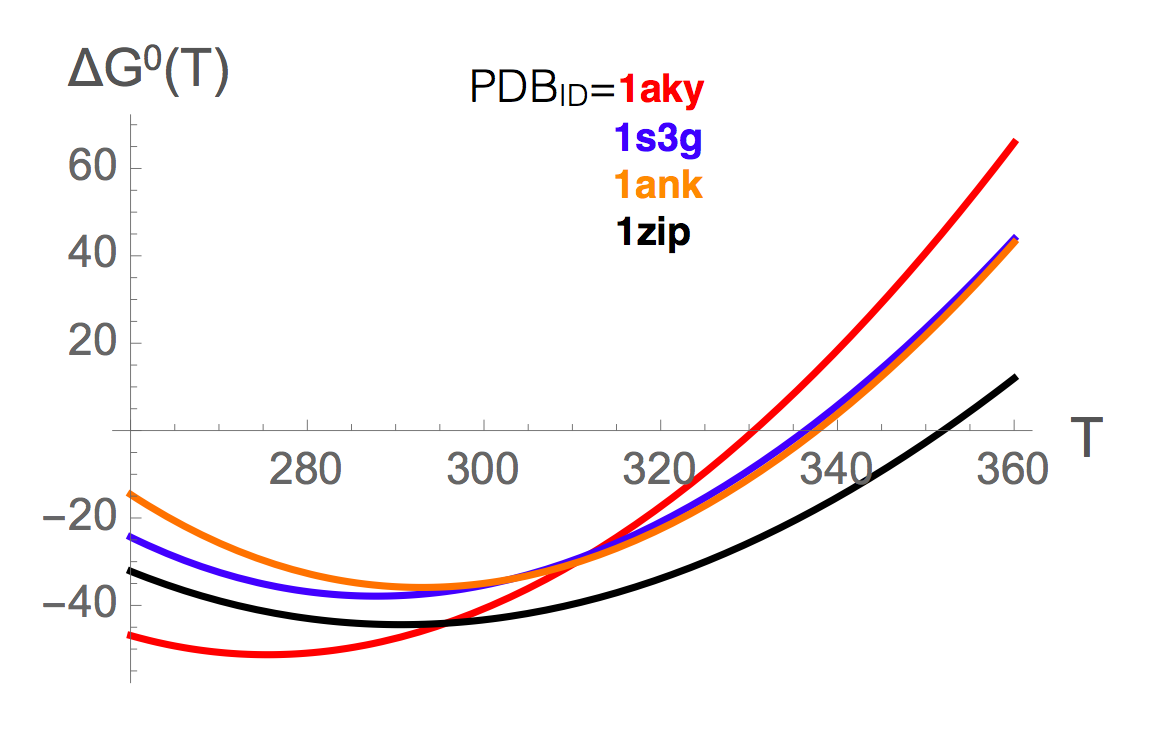}}
\caption{Stability curves of  proteins belonging to the adenylate kinase family and identified by the PDB \cite{PDBank} code 1aky (from {\it Saccharomyces cerevisiae)}, 1s3g (from {\it Sporosarcina globispora}), 1ank (from {\it Escherichia coli}) and 1zip (from {\it Geobacillus stearothermophilus}). Temperatures are measured in K and folding free energies in kJ/mol. (a) Experimental stability curves; (b) Stability curves predicted by SCooP \cite{scoop2}.} 
\label{stabcurve}
\end{center}
\end{figure}

The stability of a protein structure is defined by its standard folding free energy:
\begin{equation} \Delta  G^0 =  G^0({\rm folded\; state}) -  G^0({\rm unfolded \; state}) \quad, \end{equation}
which depends on all environmental parameters such as the solvent, pH, pressure and temperature. If we only consider the dependence on the temperature, the stability curve of a protein \cite{stab} has a inverted bell-shape (Fig. \ref{stabcurve}) and is described by the Gibbs-Helmholtz equation, which is  valid for proteins that fold according to a two-state transition: 

\begin{equation}\Delta G^0(T) = \Delta H^0_m \left( 1 - \frac{T}{T_m}\right) - \Delta C^0_P \left[(T_m - T) + T\,  \text{Log}\left(\frac{T}{T_m}\right)\right] \quad , \label{Gibb} \end{equation}
where $T_m$ is the melting temperature of the protein, {\it i.e.} the temperature at which $\Delta G^0$ vanishes,  $\Delta H^0_m$ is the standard folding enthalpy measured at $T_m$ and $\Delta C^0_P$ the folding heat capacity assumed to be $T$-independent. Note that $T_m$ is  here the denaturation temperature at high $T$; there is also a denaturation temperature at low $T$ which we  do not consider here as water usually freezes before the protein denatures.

Two complementary descriptions of  protein stability are commonly used: the thermodynamic stability is identified by $\Delta G^0(T_r)$ at room temperature ($T_r$), whereas the thermal stability is defined by the melting temperature $T_m$. 

Thermal and thermodynamic stabilities are not always correlated, which is obvious from the Gibbs-Helmholtz equation (\ref{Gibb}) when $T=T_r$. The molecular reason of the lack of correlation between the two stabilities is that the 20 amino acids are of different physicochemical types, and that their energetic contributions have different temperature dependences. 

Upon substitution of one amino acid into another, or of several amino acids, the so-called wild-type protein is transformed into a mutant protein, and this has an impact on all its biophysical properties. The folding free energy gets modified as: 
$ \Delta \Delta G^0 = \Delta G^{0\;{\rm mutant}} - \Delta G^{0\;{\rm wild}}$, 
the melting temperature as:
$ \Delta T_m= T_m^{\rm mutant} - T_m^{\rm wild}$, and similarly for the changes in folding heat capacity ($\Delta \Delta C^0_P$) and  enthalpy ($\Delta \Delta H_m^0$).
The changes in thermodynamic and thermal stabilities upon mutations are more -- but still imperfectly -- correlated than the stabilities themselves, as can be deduced from the relation:
\begin{eqnarray} 
\Delta \Delta G^0 (T_r) \simeq  && \frac{\Delta H^0_m T_r }{(T_m^{\rm wild})^2}  \Delta T_m + \Delta \Delta H^0_m \left[1 - \frac {T_r} {T_m^{\rm wild}} +  \frac {T_r} {(T_m^{\rm wild})^2}\Delta T_m \right] \nonumber \\ 
&&+  \Delta \Delta C^0_P \left[ T_r-T_m^{\rm wild}-T_r \log {\frac {T_r}{T_m^{\rm wild}}}\right]  + 
\mathcal{O} \left ( \frac{\Delta T_m^{\; 2}} {T_m^{{\rm wild}\; 2}} \right ) \qquad .
\label{DDGDTM} 
\end{eqnarray}
The linear correlation between $\Delta \Delta G^0 (T_r)$ and $ \Delta T_m$ values was found to be equal to -0.7 on about 450 experimentally characterized mutations \cite{Data}. 

\section{Statistical potentials}\label{sec:sp}

\begin{figure}[h]
\begin{center}
\subfigure[]
{\includegraphics[scale=0.62]{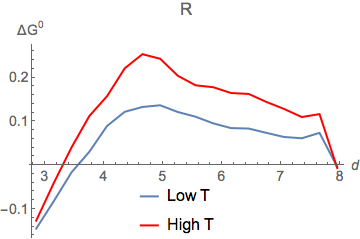}}
\subfigure[]
{\includegraphics[scale=0.62]{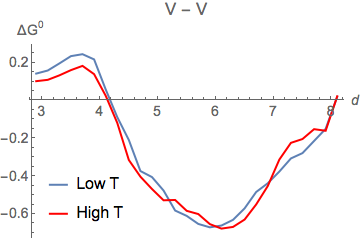}}
\caption{Examples of distance potentials obtained from sets of mesostable proteins (blue curves) and thermostable proteins (red curves). The folding free energy (in kcal/mol) is given as a function of inter-residue distance $d$ (in \AA). (a) $\Delta W(R, d, T)$ potential based on the propensity of the amino acid of type R (arginine) to be  at distance $d$ from any other amino acid. (b) $\Delta W ( V, V, d, T)$ potential based on the propensity of two amino acids of type V (valines) to be separated by a distance $d$.} 
\label{potentials}
\end{center}
\end{figure}

To estimate the energy of a protein, two types of energy functions are commonly used. The semi-empirical force fields assume an analytic form for the different types of interactions with parameters that are transferred from experiments on small molecules or from quantum chemistry calculations \cite{semi}. The statistical potentials do not assume any analytical form, and are based on a coarse-grained representation of protein structure, which is a computational advantage if they have to be applied on an omics scale or  in other time-consuming calculations. They have moreover the advantage of considering implicitly the effect of the solvent ({\it i.e.} water for soluble proteins). They are obtained by deriving a potential of mean force (PMF) from  frequencies of associations  of  structure and sequence elements  in a dataset of known 3D protein structures, using the Boltzmann law \cite{stat1,stat2}. The simplest PMF can be written as:
\begin{equation}
\Delta W(s, c) \cong  -  k T\, \ln\, \frac{F(s,c)}{F(s) F(c)} \qquad,
\label{En}
\end{equation}
where $c$ and $s$ are  structure and  sequence elements respectively, $F$ represent the relative frequencies of  $c$ and/or $s$,  and  $k$ is the Boltzmann constant. Sequence elements are single amino acids, amino acid pairs, triplets, etc, and structure elements are inter-residue distances, main chain torsion angle domains, solvent accessibility, etc.  More complex potentials can be constructed by considering more than two structure elements and/or sequence elements. Examples of such  potentials are given in Fig. \ref{potentials}.

The derivation of these potentials is based on several assumptions. All the conformations accessible to a given protein are known to follow the Boltzmann law, but here it is assumed that the  sequence-structure elements  observed in the lowest free energy conformation of the whole set of different proteins  follow the same law. Another strong assumption is that the  unfolded state may be modeled by conformations that are sequence-independent.

A property of these potentials that can be either inconvenient or interesting -- according to the issue -- is that they depend on the dataset from which they are derived. Such a  dependence can lead to unduly biased potentials  if the dataset is noisy, redundant, or unbalanced. But it is also possible to take advantage of this dependence. For example, by dividing the protein structure set into a subset of thermostable proteins ($T_m\ge 65$\textdegree C, average $\bar{T}_m \cong 80$\textdegree C, labeled by $\bigtriangleup$) and a subset of mesostable proteins  ($T_m <65$\textdegree C,  $ \bar{T}_m  \cong 50$\textdegree C, labeled by $\bigtriangledown$), and deriving the potentials separately from both subsets, we obtain $T$-dependent potentials \cite{statT}:
\begin{equation}
\Delta W(s, c, T) \cong  -  k T\, \ln\, \frac{F(s,c, T)}{F(s, T) F(c, T)} \quad ,
\label{En_T}
\end{equation}
with $T\cong \bar{T}_m$. The potentials derived from  mesostable (thermostable) proteins yield energy estimations at low (high) $T$.  We would like to emphasize that this is currently the only way of obtaining $T$-dependent free energy estimations. These potentials allowed us to get insight  into the $T$-dependences of specific interactions. For example, as shown in Fig. \ref{potentials}, the relative propensity of the positively charged residue R (arginine) to be at a distance $d$ from any other amino acid is strongly $T$-dependent, whereas the relative weight of the effective interactions between two hydrophobic residues V (valines) is $T$-independent.  
 Note also that, in principle, one could define (overlapping) sets of protein structures with increasing $T_m$-values, to get almost continuous $T$-dependence. However, this is currently impossible due to the limited number of protein structures of experimental determined $T_m$, which do not ensure reliable statistics. 

\section{Prediction of the stability curve of proteins}\label{sec:scoop}

Using the standard and the $T$-dependent statistical potentials described above, we developed a method called SCooP to predict the Gibbs-Helmholtz $\Delta G^0(T)$ curve of a target protein with known 3D structure \cite{scoop1,scoop2}. SCooP is thus able to predict all the thermodynamic quantities that characterize the folding transition, namely the melting temperature $T_m$ as well as the folding heat capacity $\Delta C^0_P$, enthalpy $\Delta H^0_m$ and free energy  $\Delta G^0(T_r)$ at room temperature. 

These thermodynamic descriptors were obtained from specific combinations of the statistical potentials described in Eqs (\ref{En},\ref{En_T}) with other protein characteristics such as their number of residues $N$ and their total  surface area  $ASA$. In particular, the functional form for the folding heat capacity $\Delta C_P^0$ was expressed as:
\begin{equation}
\Delta C_P^0 = \alpha_0 N +  \frac{\alpha_1 N + \beta_1}{N} ASA + \sum_{\nu=2}^5 \frac{\alpha_\nu N+ \beta_\nu}{N} \left( \Delta W_\nu^{\bigtriangleup} - \Delta W_\nu^{\bigtriangledown} \right) \qquad ,
\label{CpI}\end{equation}
where $\alpha_\nu$ and $\beta_\nu$ are parameters to be optimized and the $T$-dependent potentials $ \Delta W_\nu^{\bigtriangleup / \bigtriangledown}$ are defined 
in Eq.(\ref{En_T})\footnote{The following combinations of potentials were used: 
$\Delta W_2(T) = [\Delta W(a,t,T) + \Delta W(a,a,t,T)]$, $\Delta W_3(T) = [\Delta W(a,\mathcal{ A},T) + \Delta W(a,a,\mathcal{ A},T)]$,   
$\Delta W_4(T) = [\Delta W(a,d,T) + \Delta W(a,a,d,T)]$, and $\Delta W_5(T) = [\Delta W(a,\mathcal{ A},d,T) + \Delta W(a,a,\mathcal{ A},\mathcal{ A},d,T)]$, where $a$ stands for  amino acid type, $d$ for  inter-residue distance, $t$ for  main chain torsion angle domain, and $\mathcal{ A}$ for solvent accessibility.}. 
Analogous expressions with slightly different combinations of features were used for $T_m$ and $\Delta H_m^0$  \cite{scoop2}. These functional forms were chosen on the basis of  biophysical insights and  empirical evidence. For example, we considered the first two terms in the right hand side of expression (\ref{CpI}) on the basis of the experimental observation that $\Delta C_P$, $N$ and $ASA$ are quite well correlated. The last terms in  (\ref{CpI}) come  from the $\Delta C_P$ definition : 
\begin{equation}
\Delta C_P^0 =  C_P^0 ({\rm folded}) - C_P^0 ({\rm unfolded}) =\frac{\partial \Delta H^0(T)}{\partial T} \simeq  \frac{\partial \sum_\nu \gamma_\nu (N) \Delta W_\nu(T)}{\partial T}  \qquad ,
\label{CpII}\end{equation} 
and the approximation consisting of expressing the folding entalphy $\Delta H^0 (T)$ as a specific combination of the statistical potentials at  temperature $T$ weighted with parameters that depend on the number of residues  $\gamma_\nu (N) =  \alpha_\nu  + \frac{\beta_\nu}{N}$. Since the continuous $T$-dependence of the potentials is not known, we  replaced the derivative by the finite difference between thermo- and mesostable potentials: 
\begin{eqnarray}
 \frac{\partial \sum_\nu \gamma_\nu (N) \Delta W_\nu(T)}{\partial T}  \simeq  \sum_\nu\frac{\gamma_\nu (N)}{\Delta T} \left( \Delta W_\nu^{\bigtriangleup} -  \Delta W_\nu^{\bigtriangledown} \right) \qquad ,
\label{CpIII}\end{eqnarray} 
where $\Delta T$ is a constant that was integrated in the parameters.

All the parameters that appear in the functional form of $\Delta C_P^0$ were  optimized by minimizing the mean square error between the predicted and the experimental quantities of  $\ell \approx 220$ proteins with known 3D structure and stability curve:
$ \sigma^2 = \frac{1}{\ell}\sum_{k=1}^{\ell} \left(\Delta C_{P \; k}^{0\; pred}- \Delta C_{P \; k}^{0\; exp}\right)^2 $.
The same procedure was used to identify the parameters appearing in the functional forms of $T_m$  and $\Delta H_m^0$ (not shown). The predicted values of $\Delta C_P^0$, $T_m$  and $\Delta H_m^0$ were then used to draw the full stability curve using the Gibbs-Helmholtz equation (\ref{Gibb}). The SCooP predictor is freely available on the website http://babylone.ulb.ac.be/SCooP.

One of the strengths of the SCooP method is the computation speed of the full $T$-dependence of the folding free energy, which allows large-scale analyses of the entire structurome. Another of its strengths is its good performance: the linear correlation coefficients between the experimental and the predicted values of  $T_m$, $\Delta C^0_P$ and $\Delta H^0_m$ for the dataset of about 220 proteins are all three equal to  0.7, as computed with a leave-one-out cross validation procedure. The scores increase up to  0.8 after the exclusion of 10\% outliers. Examples of stability curves predicted by SCooP  for  proteins belonging to the adenylate kinase family are given in Fig. \ref{stabcurve}b. As seen from the comparison with Fig. \ref{stabcurve}a  in which the experimentally determined stability curves are plotted, the SCooP predictions reproduce rather well the full  stability curves. 

\section{Prediction of the stability changes upon mutations}\label{sec:pophot}
\subsection{Thermodynamic  stability}

It can be expected to be easier to predict changes in  stability upon amino acid substitutions than the actual stability of proteins. Indeed, the large majority of single-site mutations do not affect the structure of the protein, but simply modify its stability, solubility, affinity for biomolecular partners or function. The PoPMuSiC program \cite{PoP1,PoP2} , which is available  on 
http://www.dezyme.com, predicts changes in the thermodynamic stability of a protein on the basis of its 3D structure and statistical potentials.
The $\Delta\Delta G^0(T_r)$ of a point mutation is expressed as the sum of 16  terms :
\begin{equation}
\Delta \Delta G^0(T_r) =\sum_{\nu=1}^{13} \alpha_\nu(\mathcal{ A}) \Delta \Delta W_\nu + \alpha_{+}(\mathcal{ A}) \Delta V_+ + \alpha_{-}(\mathcal{ A}) \Delta V_- + \alpha_{I}(\mathcal{ A})  \qquad,\label{pop}
\end{equation}
where the 13  $\Delta \Delta W_\nu$  terms are  changes in folding free energy upon mutation computed with 13 different  statistical potentials $\Delta W$ defined in Eq.(\ref{En})\footnote{ These are: 
$\Delta W(a,d)$, $\Delta W(a,t)$, $\Delta W(a,\mathcal{ A})$, $\Delta W(a,a,d)$,  $\Delta W(a,a,t)$, $\Delta W(a,t,t)$, $\Delta W(a,\mathcal{ A},\mathcal{ A})$,  $\Delta W(a,t,d)$, $\Delta W(a,\mathcal{ A},d)$,   
$\Delta W(a,a,\mathcal{ A})$, $\Delta W(a,t,\mathcal{ A})$, $\Delta W(a,\mathcal{ A},a,\mathcal{ A},d)$, $\Delta W(a,t,a,t,d)$.}.
The coefficients $\alpha_\nu$ are taken as sigmoid functions of the solvent accessibility of the mutated residue:

\begin{equation}\alpha_\nu(\mathcal{ A})=\omega_\nu [1+exp^{-r_\nu(\mathcal{ A} - b_\nu)}]^{-1}+\phi_\nu,\end{equation}
where $\omega_\nu$, $r_\nu$, $b_\nu$ and $\phi_\nu$ $\in \mathbb{R}$ are parameters to be optimized.  
 This $\mathcal{ A}$-dependence is justified by the fact that the weight of the interactions vary continuously from the protein core to the surface. The  $\Delta V_\pm$ are volume terms defined by
$\Delta V_\pm = \theta ( \pm \Delta V ) \| \Delta V\|$,
where $\theta(V)$ is the Heaviside function. They are associated with the difference in volume of the mutant 
and wild-type amino acids and provide a description of the impact of the creation of a cavity or the accommodation of stress inside the protein structure. 

\begin{figure}
\centering
\subfigure[]
{\includegraphics[scale=0.21]{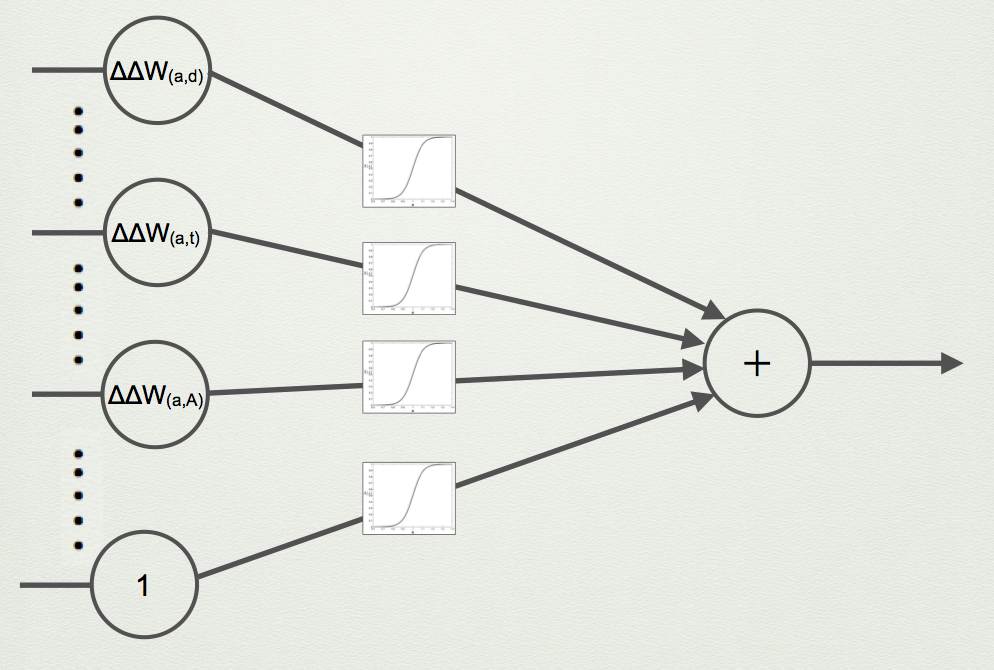}}
\subfigure[]
{\includegraphics[scale=0.197]{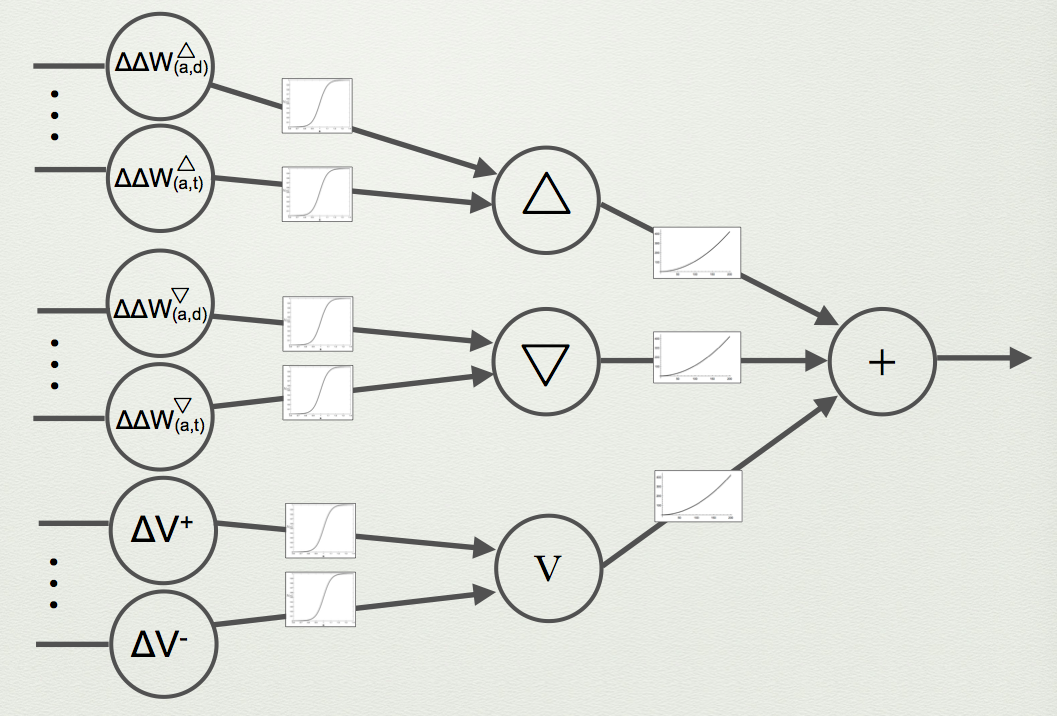}}
\caption{Schematic representation of the ANNs used for the parameter identifications. (a) PoPMuSiC and HoTMuSiC: 2-layer
ANN (perceptron) with sigmoid activation functions and input neurons encoding $T$-independent
potentials, volume terms and an independent term; (b)  Tm-HoTMuSiC: 3-layer ANN,
consisting of 3 perceptrons with sigmoid weights; the neurons of the first (second) perceptron encode the high-$T$ (low-$T$) potentials, and those of the third perceptron the volume and independent terms. The outputs of these three perceptrons are the
inputs of another perceptron with polynomial weight functions.} 
\label{ann}
\end{figure}

The optimal ($\omega_\nu$, $r_\nu$, $b_\nu$, $\phi_\nu$) parameter values were determined on the basis a  dataset of $\ell=2648$ single-site  mutations in 131 different proteins with experimentally determined 3D structure and $\Delta\Delta G^0(T_r)$. For that purpose, an artificial feedforward neural network (ANN)  was used, as shown in Fig. \ref{ann}.a, with the mean square error between the experimental and predicted folding free energy changes upon mutation  as cost function :
$\sigma^2 = \frac{1}{\ell}\sum_{k=1}^{\ell} \left(\Delta \Delta G_k^{0 \; {\rm pred}}(T_r)- \Delta \Delta G_k^{0 \; {\rm exp}}(T_r)\right)^2$.
The  parameter identification was particularly complex due to the size and roughness of the parameter space, which makes the search for the global minimum complicated and yields  a high probability of getting stuck in local minima. Therefore, the initial values of the weights were chosen randomly,  the initialization and  training processes were repeated  30 times,  and the solution reaching the lowest $\sigma$-value was chosen. 

The performance of PoPMuSiC was  computed in 5-fold cross validation. The linear correlation coefficient between experimental and predicted $\Delta \Delta G^0(T_r)$ values is equal to $r=0.6$ and increases to 0.8 with 10\% outliers suppressed; the mean square error $\sigma$ is equal to 1.1 kcal/mol and decreases to 0.9 kcal/mol without 10\% outliers. This score is comparable or better than that of other approaches. It must be emphasized  that PoPMuSiC is extremely fast, and predicts the $\Delta \Delta G^0(T_r)$ of all possible point mutations in a medium-size protein in a few minutes. It can thus be applied on a large, structuromics, scale. Of course this high speed comes at the expense of some accuracy. 

\begin{figure}[h]
\begin{center}
\includegraphics[scale=0.32]{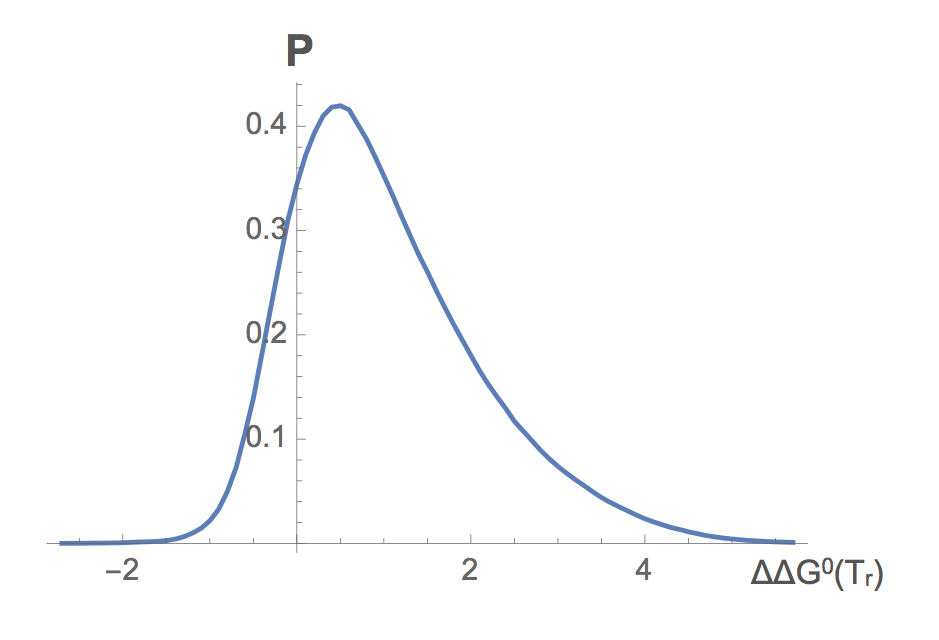}
\caption{Distribution of the $\Delta \Delta G^0(T_r)$ values for all possible amino acid substitutions in a set of 15 protein structures predicted  by PoPMuSIC \cite{PoP1,PoP2}.} 
\label{distri}
\end{center}
\end{figure}

The distribution of $\Delta \Delta G^0(T_r)$ values computed by PoPMuSiC for all possible amino acid substitutions (about 10$^5$ mutations in total) in a set of 15 protein structures is shown in Fig. \ref{distri} \cite{ifac}. The large majority of the mutations have  positive $\Delta \Delta G^0(T_r)$ values  and  are thus destabilizing with our conventions. This is of course expected, as protein sequences are relatively well optimized for the stability of the native fold. Note, however, that natural proteins are never perfectly optimal, because some residues are optimized for function rather than for stability, but also, probably, because natural evolution has not had the time to reach optimality. 

\subsection{Thermal stability}

As thermodynamic and thermal stabilities are far from perfectly correlated, as clear from Eq.(\ref{DDGDTM}), it is necessary to develop methods that predict directly $\Delta T_m$ values, instead of  approximating them on the basis of $\Delta \Delta G^0(T_r)$ predictions. This is the goal of  HoTMuSiC~\cite{HoT}, which is available on http://www.dezyme.com. The first version of this program uses the relation:
\begin{equation}
\Delta T_m^{\text{HoT}} = \frac{1}{a N +c}  \left( \sum_{\nu=1}^{9} \alpha_{\nu}(\mathcal{A})  \Delta\Delta W_\nu  + \alpha_{+}(\mathcal{A}) \Delta V_+ + \alpha_{-}(\mathcal{A}) \Delta V_- + \alpha_{I}(\mathcal{A}) \right), \qquad
\label{hot}
\end{equation}
where $N$ is the number of residues in  the protein, $a$, $c$ $\in \mathbb{R}$, and the coefficients $\alpha_\nu(\mathcal{A})$ are sigmoid functions of $\mathcal{A}$. This equation is inspired by Eq.(\ref{DDGDTM}) with all but the first term on the right hand side overlooked, and is thus very similar to Eq.(\ref{pop}) for $\Delta \Delta G^0(T_r)$. The 9 $\Delta\Delta W_\nu$ terms are computed from the first 9  of the 13 statistical potentials used in Eq.(\ref{pop}). The optimal parameter values appearing in this equation were identified on the basis of $\ell=1531$ single-site  mutations with known 3D structure and $\Delta T_m$, the cost function $\sigma^2 = \frac{1}{\ell}\sum_{k=1}^{\ell} \left(\Delta T_m^{\rm pred}- \Delta T_m^{\rm exp}\right)^2$, and a feedforward ANN (Fig. \ref{ann}a).

The second HoTMuSiC version  is inspired by  the full Eq.(\ref{DDGDTM}), and uses the $T$-dependent statistical potentials of Eq.(\ref{En_T}); it requires as input, in addition to the protein structure, the melting temperature $T_m^{\rm wild}$ of the wild-type protein:

\begin{eqnarray}
\Delta T_m^{\text{T$_{\rm m}$HoT}} = &&\beta^{\triangle}(T_m^{\rm wild},N)\sum_{\nu=1}^{5} \alpha_{\nu}^{\triangle}(\mathcal{A}) \Delta \Delta W_\nu^{\triangle} + \beta^{\bigtriangledown}(T_m^{\rm wild},N) \sum_{\nu=1}^{5} \alpha_{\nu}^{\bigtriangledown}(\mathcal{A}) \Delta \Delta W_\nu^\bigtriangledown  \nonumber \\
&&+  \beta^{V}(T_m^{\rm wild},N) \left[ \alpha_{+}(\mathcal{A}) \Delta V_+ +\alpha_{-}(\mathcal{A}) \Delta V_- +\alpha_{I}(\mathcal{A})\right] \qquad . \label{hotp}
\end{eqnarray}
The  5 $\Delta \Delta W_\nu^{\triangle}$  ($\Delta \Delta W_\nu^{\bigtriangledown}$) terms are folding free energy changes computed from the dataset of thermostable (mesostable) proteins using Eq.(\ref{En_T})\footnote{The following combinations were used: $\Delta W(a, t, d, T)$, $[\Delta W(a,d,T)+\Delta W(a,d,a,T)]$, 
$[\Delta W(a, \mathcal{A}, d, T) +\Delta W(a, \mathcal{A},d, s, \mathcal{A},T)]$,  
$[\Delta W(a, \mathcal{A},T)+\frac 12  \Delta W(a, \mathcal{A},\mathcal{A},T)+\frac 12   \Delta W(a,a,\mathcal{A},T)]$,
$[\Delta W(a,t,T)  + \frac 12 W(a,t,t,T)  +\frac 12 W(a,a,t,T)] $}. 
$\beta^{\triangle}$, $\beta^{\bigtriangledown}$ and $\beta^V$ 
are polynomial functions of $T_m^{\rm wild}$  and the number of residues $N$. Their functional form guessed from  Eq.(\ref{DDGDTM}) is:
\begin{equation}
\beta^I(T_m^{\rm wild},N) = \frac{\gamma^I (T_m^{\rm wild})^2 + \delta^I T_m^{\rm wild} + \xi^I}{a^I N+ c^I} , \label{ACII}\end{equation}
with  $I$=$\triangle$, $\bigtriangledown$ or $V$. The dependence on the number of residues $N$ comes from the enthalpy factor $\Delta H_m$  in Eq.(\ref{DDGDTM}), as these two quantities show a good correlation.

To identify the parameters of this second method, a 3-layer ANN is used which is shown schematically in Fig. \ref{ann}b. The input layer consists of three sets of neurons that encode the mesostable potentials, the thermostable potentials, and the volume and independent terms, respectively. These three sets of  neurons are the inputs of three perceptrons, whose outputs are the three neurons of the hidden layer. These are the input neurons of yet another perceptron, which yields a $\Delta T_m$-prediction as output. This predicted $\Delta T_m$ is finally  averaged with the $\Delta T_m$ predicted by the first HoTMuSiC method (Eq.(\ref{hot})). The initialization and identification procedures of all parameters and  the cross validation procedure are the same as for the first method. 

The root mean square error between experimental and predicted $\Delta T_m$-values is equal to  $\sigma=$ 4.6\textdegree C for HoTMuSiC and $\sigma=$ 4.5\textdegree C for $T_m$-HoTMuSiC;  the associated Pearson correlation coefficients $r$ are both equal to  0.6.  When  10\% outliers are excluded, $\sigma$  decreases to 3.3 and 3.2\textdegree C  and $r$ rises to 0.7.

Finally note that, because of the two volume terms and the energy-independent term in Eqs (\ref{pop},\ref{hot},\ref{hotp}), the symmetry:

\begin{equation}\Delta T_m({\rm wild} \rightarrow {\rm mutant})= - \Delta T_m({\rm mutant} \rightarrow {\rm wild}) \label{symm}\end{equation} 
is explicitly broken, and similarly for $\Delta \Delta G^0(T_r)$ \cite{ifac}. Due to these symmetry-breaking  terms and the fact that the large majority of experimentally characterized mutations in the training dataset are destabilizing, the predictions of $\Delta \Delta G^0(T_r)$ and $\Delta T_m$ are on the average more accurate for  destabilizing mutations,  at the price of a decrease in the prediction performance of  stabilizing mutations. But actually, the stabilizing mutations  are the most interesting ones in most protein design and optimization applications. To solve this problem and achieve higher prediction scores for stabilizing mutations, we are currently developing new PoPMuSiC and HoTMuSiC versions that preserve the symmetry of Eq.(\ref{symm}). 

\section{Application to a conformational disease protein}
To illustrate the power of the above described approaches, we describe one of their successful applications in the context of conformational diseases \cite{Ataxin}. Such disorders are caused by the misfolding of  specific proteins, which become inactive or toxic \cite{confdisease1, confdisease2}. Often these proteins form soluble or insoluble aggregates. Creutzfeldt-Jakob and Alzheimer  are  well known examples of this class of diseases. Misfolded conformations correspond to  free energy minima that differ from those of  the folded and unfolded states. They may be thermodynamically or kinetically favored under certain conditions or for certain protein variants. The precise mechanism by which misfolding is initiated and propagates  to other proteins is only partially known. 

Spinocerebellar ataxia type 3  is a less known example of conformational disease \cite{polyQ1,polyQ2}. It is characterized by progressive neuronal dysfunction and the presence of neuronal inclusions
containing fibrillar aggregates of ataxin-3 proteins. Ataxin-3 contains a globular, structured domain, called  Josephin, and a flexible polyglutamine (polyQ) domain whose  repeat-length
modulates pathogenicity. It has been suggested  that the fibrillogenesis pathway of ataxin-3 starts with a polyQ-independent step mediated by Josephin domain interactions, followed by a polyQ-dependent
step. To test the involvement of the Josephin domain in ataxin-3 fibrillogenesis, the isolated Josephin domain was in turn stabilized and destabilized, and the consequences on the aggregation tendencies of this domain taken individually and of the full ataxin-3 protein were analyzed. Two ataxin-3 variants were considered, a non-pathological  variant with a 15-residue-long polyQ tract (ataxin-3(Q15)), and a pathological variant with a 64-glutamine tract (ataxin-3(Q64)) \cite{Ataxin}.

\begin{table}[h]
\begin{center}
\begin{tabular}{|c|c|c|c|c|c|c|c|}
\hline
\textbf{ Mutants }     & \textbf{Prediction }  & \multicolumn{6}{|c|}{\textbf{Experiment}}  \\ 
                    &                    &\multicolumn{2}{|c|}{Josephin}&\multicolumn{2}{|c|}{Ataxin-3(Q15)}&\multicolumn{2}{|c|}{Ataxin-3(Q64)}\\
                    &                   &  $T_m$   (\textdegree C) & Agg (h)&  $T_m$   (\textdegree C) & Agg (h)&  $T_m$   (\textdegree C) & Agg (h) \\ 
\hline   
wild-type     &       -           &   51.3  & 79.4    & 49.2 & 24.0  & 51.0 & 11.7   \\ 
  \hline                  
R103G        &                    & 55.9 & 147.5   & 52.9 & 48.9 & 52.1 & 25.3   \\ 
S81A          &   stabilizing   &  54.1 & 115.2   & 50.6 & 32.2 & 50.8 & 26.3   \\ 
R103G/S81A &                  &  55.4 & \textgreater 150  &53.4 & 50.0  &52.4 & 40.6   \\ 
  \hline  
L169H         &   destabilizing & 47.5   &  4.0  & - & - & - & -    \\ 
   \hline  
\end{tabular}
 \vspace{0.3cm}
\caption{Predicted stabilizing and destabilizing mutations in the Josephin domain and experimental characterization of the thermal denaturation and aggregation midpoints ($T_m$ and Agg) of the Josephin domain and two ataxin-3 variants of different polyQ length (Q15 and Q64). No results are shown for the L169H mutant in the two ataxin-3 variants because soluble expression was not achieved.  
\label{josephin}}
\end{center}
\end{table}

To modify the stability of the Josephin domain, the PoPMuSiC software was used.
Two mutations were selected, which  were predicted as the most stabilizing and are located sufficiently far from the active site to avoid affecting the protein's function: R103G and S81A\footnote{The first letter denotes the wild-type amino acid and the second the mutant; the number indicates the position in the sequence}. Furthermore, four destabilizing mutations were chosen in silico, but only one resulted in soluble protein expression: L169H. The $T_m$ and aggregation propensities of the wild-type and of these 3 protein mutants, as well as of the double mutant R103G/S81A, were experimentally measured. As shown in Table \ref{josephin}, the mutations predicted to stabilize the Josephin domain actually do so. They also tend to stabilize the complete ataxin-3 protein. Moreover, these mutants  significantly decrease the aggregation propensities: the aggregation midpoint is increased both for the Josephin domain and the two ataxin-3 variants Q15 and Q64. Moreover, the mutation predicted to destabilize the Josephin domain actually does so and moreover drastically decreases the aggregation midpoint. 

These results show that changing the thermodynamic stability of the Josephin domain modulates ataxin-3 fibrillogenesis, and support the hypothesis that the first stage of ataxin-3 fibrillogenesis is
caused by interactions involving the non-polyQ containing Josephin domain. They also show that the use of in silico predictors such as PoPMuSiC is very helpful to detect very rapidly  relevant candidate mutations, which can be exploited to optimize 	a protein system or to get insights into  biological or biomedical issues.

\section*{Acknowledgments}
This paper reminded   of those  unforgettable times spent by one of us (MR) with Philippe Spindel  and Fran\c cois Englert during the early days of 11D supergravity -- and later with some strange mathematician who  insisted on drawing the time axis horizontally. Philippe, thanks for all.

\providecommand{\href}[2]{#2}\begingroup\raggedright
\end{document}